# Benchmarking Over The Grid Using SEE Virtual Organization.

January 2007

## Technical Report

**Ioannis Kouvakis[1] and Fotis Georgatos[2]**

[1] **Student of Univerity of Aegean, Department of Mathematics,**
gkouvakis@hep.ntua.gr
http://www.kouvakis.gr/projects.html
(special Thanks to my thesis advisor Dr. Ioanis Gialas for giving me the opportunity to do research on Grid)

[2] **University of Cyprus, Department of Informatics,**
fotis@mail.cern.ch

# Introduction

Grids include heterogeneous resources, which are based on different hardware and software architectures or components. In correspondence with this diversity of the infrastructure, the execution time of any single job, as well as the total grid performance can both be affected substantially, which can be demonstrated by measurements.

# Approach

In order to see this heterogeneity of the grid we send a simple script job over some sites of the Grid using the SEE V.O. The script had 2 parts. The first part was taking information from the worker node that the script was running, such as Kernel Version, Linux Distribution, Environment Variables, Packages and Kernel Messages and also Hardware Information concerning Cpu Model / Vendor / Speed, Memory Size, Hard Disks and other Media Details. The second part of the script was downloading / compiling and running the Lmbench Benchmarking Suite that it was analyzing the performance of the current node.

The job was send several times over different worker nodes of each site. Based on the fact that we have homogeneity inside each site, which it was true in almost all the sites, we were able to get some statistical information and compare those sites.

# Results And Comparison Tables From LMBench.

All the results from the script and the lmbench can be downloaded from http://www.kouvakis.gr/lang_en/projects/grid.html . Also you can download the script and excel worksheets that were used to get all the pies and the statistical tables.

The information we got from the first part of the script, helped us make the following pies concerning cpu models that a job can run into and the numbers of cpu each worker node has. That information is per Node meaning that the total numbers of worker nodes were added and each slice is the percentage of those nodes that using this model of processor or cpu numbers. We are pleased to see that a great percentage use 4 cpu hyper threaded or not. We also see only a small number of nodes are based on AMD processors.

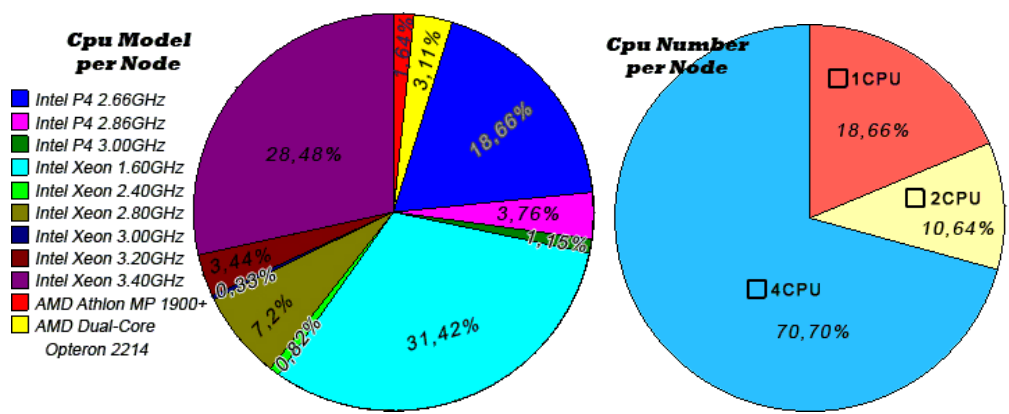

Also we can see the cpu model per site, meaning that after counting the total sites what percentage of them use each model. In the Cpu Speed per cpus (in GHz) pie we can see the percentage of the speed that each job can theoretically use.

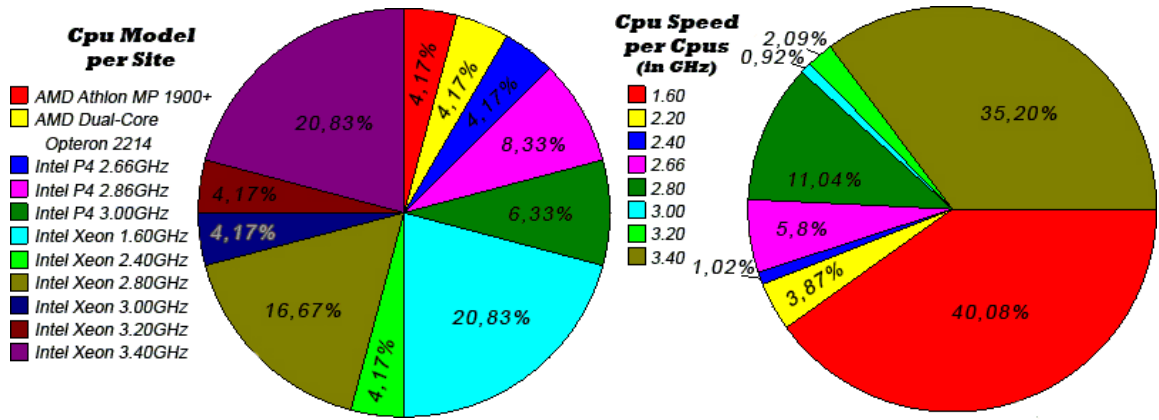

Based on the total cpu of each site from GSTAT website we were able to make those pies concerning each job that running on a worker node how much system memory can consume and also the total size of the swap memory of that node. We have a major percentage of 4096MB and 2048. The swap memory is very confusing if we consider that not all the nodes are following the 2:1 Swap/Ram rule.

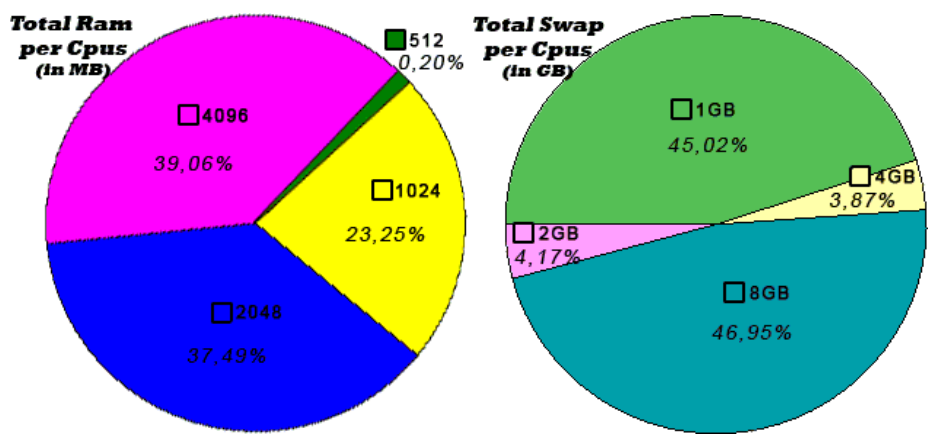

The following pies are showing the Linux Distribution and kernel version per site. The majority of the sites have Latest linux Kernel and a great percentage use take advantage of the smp technology of the cpu using the smp kernel version.

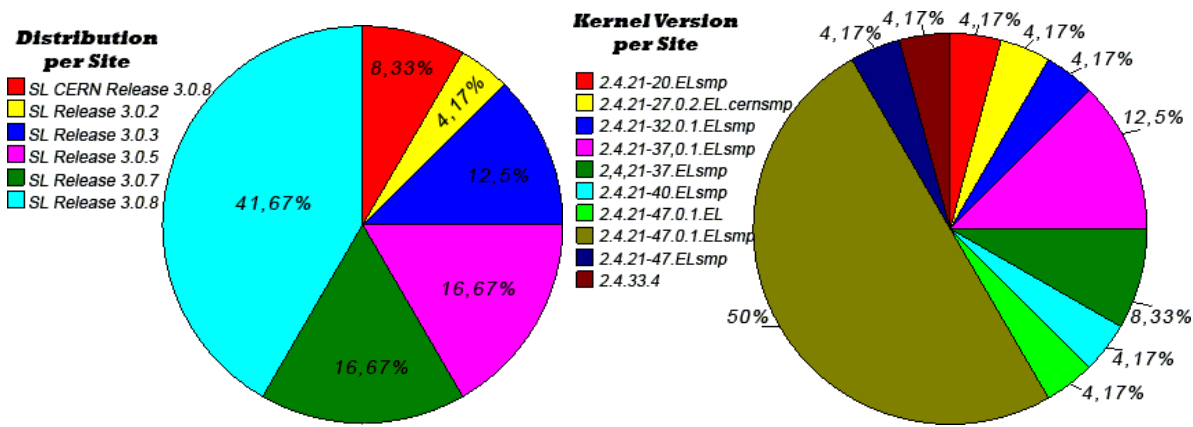

Final the following two tables are showing the number of users for those sites for known VOs and a combined with GSTAT information about hard disks.

| Site Name (short) | alice | atlas | biomed | cms | compchem | dteam | esr | gridcc | hgdemo | lhcb | magic | ops | planck | see | seegrid |
|---|---|---|---|---|---|---|---|---|---|---|---|---|---|---|---|
| ce.hep.ntua.gr | 200 | 200 | 50 | 202 | 0 | 51 | 0 | 0 | 0 | 200 | 0 | 0 | 0 | 0 | 0 |
| ce.phy.bg.ac.yu | 0 | 202 | 0 | 202 | 0 | 202 | 202 | 0 | 0 | 0 | 0 | 202 | 0 | 202 | 202 |
| ce.ulakbim.gov.tr (lcg) | 0 | 52 | 0 | 52 | 0 | 52 | 0 | 0 | 0 | 52 | 0 | 52 | 0 | 22 | 23 |
| ce001.imbm.bas.bg | 52 | 52 | 52 | 52 | 0 | 52 | 52 | 0 | 0 | 52 | 0 | 52 | 0 | 53 | 51 |
| ce001.ipp.acad.bg | 52 | 52 | 52 | 52 | 0 | 52 | 52 | 0 | 0 | 52 | 52 | 52 | 0 | 52 | 52 |
| ce01.afroditi.hellasgrid.gr | 202 | 202 | 202 | 202 | 202 | 202 | 202 | 0 | 202 | 202 | 202 | 202 | 202 | 202 | 0 |
| ce01.ariagni.hellasgrid.gr | 200 | 200 | 51 | 202 | 0 | 51 | 100 | 0 | 100 | 200 | 0 | 51 | 100 | 50 | 0 |
| ce.athena.hellasgrid.gr | 201 | 201 | 201 | 199 | 200 | 201 | 200 | 200 | 200 | 202 | 0 | 201 | 200 | 200 | 0 |
| ce01.isabella.grnet.gr | 201 | 201 | 201 | 199 | 200 | 201 | 200 | 200 | 200 | 202 | 0 | 201 | 200 | 200 | 201 |
| ce01.kallisto.hellasgrid.gr | 201 | 201 | 201 | 199 | 200 | 201 | 200 | 0 | 200 | 201 | 0 | 201 | 200 | 200 | 0 |
| ce01.marie.hellasgrid.gr | 101 | 101 | 101 | 103 | 101 | 101 | 101 | 0 | 101 | 101 | 101 | 101 | 101 | 101 | 0 |
| ce02.marie.hellasgrid.gr | 101 | 101 | 101 | 103 | 101 | 101 | 101 | 100 | 101 | 101 | 101 | 101 | 101 | 101 | 0 |
| ce101.grid.ucy.ac.cy | 66 | 66 | 66 | 66 | 66 | 66 | 0 | 0 | 0 | 66 | 0 | 66 | 0 | 66 | 0 |
| ctb31.gridctb.uoa.gr | 0 | 0 | 0 | 0 | 0 | 52 | 0 | 0 | 0 | 0 | 0 | 52 | 0 | 201 | 0 |
| cox01.grid.metu.edu.tr | 0 | 52 | 51 | 52 | 0 | 52 | 0 | 0 | 0 | 52 | 0 | 52 | 0 | 22 | 23 |
| grid-ce.ii.edu.mk | 201 | 201 | 201 | 199 | 200 | 201 | 200 | 200 | 200 | 202 | 0 | 0 | 200 | 200 | 201 |
| grid001.ics.forth.gr | 200 | 200 | 50 | 202 | 0 | 51 | 0 | 0 | 20 | 200 | 0 | 0 | 0 | 50 | 0 |
| grid01.cu.edu.tr | 0 | 52 | 51 | 52 | 0 | 52 | 0 | 0 | 0 | 52 | 0 | 52 | 0 | 22 | 23 |
| grid01.erciyes.edu.tr | 0 | 52 | 51 | 52 | 0 | 52 | 0 | 0 | 0 | 52 | 0 | 52 | 0 | 22 | 23 |
| kalkan.ulakbim.gov.tr | 0 | 52 | 51 | 52 | 0 | 52 | 0 | 0 | 0 | 52 | 0 | 52 | 0 | 22 | 23 |
| node001.grid.auth.gr | 201 | 201 | 51 | 201 | 0 | 52 | 0 | 0 | 100 | 201 | 0 | 51 | 0 | 51 | 0 |
| paugrid1.pamukkale.edu.tr | 0 | 52 | 51 | 52 | 0 | 52 | 0 | 0 | 0 | 52 | 0 | 52 | 0 | 22 | 0 |
| testbed001.grid.ici.ro | 201 | 0 | 201 | 0 | 0 | 201 | 201 | 0 | 0 | 201 | 0 | 0 | 0 | 201 | 201 |
| wipp-ce.weizmann.ac.il | 200 | 200 | 0 | 202 | 0 | 51 | 0 | 0 | 0 | 200 | 0 | 0 | 0 | 51 | 0 |

| Site Name | Details (Local WN storage) | | | | GSTAT | | |
|---|---|---|---|---|---|---|---|
| | Type | DF 1K-Blocks | DF Available | Details | SE Name | Av T.B. | Used T.B. |
| ce.hep.ntua.gr | Local-SATA | 76,667,520 | 61,562,168 | Maxtor 6Y080M0 SATA, 150 MB/s, 7200 RPM, 8MB | se.hep.ntua.gr | 0.77 | 0.07 |
| ce.phy.bg.ac.yu (lcg) | NFS | 118,169,880 | 24,574,584 | se.phy.bg.ac.yu | se.phy.bg.ac.yu | 0.04 | 0.09 |
| ce.ulakbim.gov.tr | NFS | 1,153,395,032 | 597,693,568 | se.ulakbin.gov.tr | se.ulakbin.gov.tr | 3.78 | 2.33 |
| ce001.imbm.bas.bg | NFS | 36,392,656 | 16,995,024 | se001.imbm.bas.bg | se001.imbm.bas.bg | 0.03 | 0.03 |
| ce001.ipp.acad.bg (glite) | Local-SATA | 116,658,284 | 100,414,408 | IBM Hitachi Deskstar 160GB SATA U150 7200RPM 8MB Buffer | se001.grid.bas.bg | 0.88 | 0.11 |
| ce002.ipp.acad.bg (lcg) | Local-SATA | 100,790,004 | 91,996,132 | IBM Hitachi Deskstar 160GB SATA U150 7200RPM 8MB Buffer | se001.grid.bas.bg | 0.88 | 0.11 |
| ce01.afroditi.hellasgrid.gr | GPFS | 629,147,136 | 552,160,768 | | se01.afroditi.hellasgrid.gr | 2.34 | 0.49 |
| ce01.ariagni.hellasgrid.gr | GPFS | 614,400,000 | 521,695,488 | | se01.ariagni.hellasgrid.gr | 2.56 | 0.35 |
| ce01.athena.hellasgrid.gr (long) | GPFS | 1,171,872,256 | 1,107,568,128 | | se01.athena.hellasgrid.gr | 9.36 | 0.01 |
| ce01.athena.hellasgrid.gr (see) | GPFS | 1,171,872,256 | 1,107,568,384 | | se01.athena.hellasgrid.gr | 9.36 | 0.01 |
| ce01.isabella.grnet.gr | GPFS | 1,714,194,432 | 1,530,016,512 | | se01.isabella.grnet.gr | 1.68 | 3.1 |
| ce01.kallisto.hellasgrid.gr | GPFS | 585,936,128 | 476,046,592 | | se01.kallisto.hellasgrid.gr | 2.45 | 0.17 |
| ce01.marie.hellasgrid.gr | GPFS | 614,401,536 | 525,807,616 | | se01.marie.hellasgrid.gr | 2.66 | 0.19 |
| ce02.athena.hellasgrid.gr (long) | GPFS | 1,171,872,256 | 1,107,568,128 | | se01.athena.hellasgrid.gr | 9.36 | 0.01 |
| ce02.athena.hellasgrid.gr (see) | GPFS | 1,171,872,256 | 1,107,568,384 | | se01.athena.hellasgrid.gr | 9.36 | 0.01 |
| ce02.marie.hellasgrid.gr | NFS | 196,015,808 | 129,851,608 | se01.marie.hellasgrid.gr | se01.marie.hellasgrid.gr | 0.16 | 0.01 |
| ce101.grid.ucy.ac.cy | Local-SATA | 10,317,828 | 9,681,620 | West. Dig. WD800JD SATA 80 GB, 300 MB/s, 8 MB, 7200 RPM | se101.grid.ucy.ac.cy | 2.59 | 0.59 |
| ctb31.gridctb.uoa.gr | Local-SATA | 18,975,324 | 16,319,624 | West. Dig. WD800JD SATA 80 GB, 300 MB/s, 8 MB, 7200 RPM | | | |
| cox01.grid.metu.edu.tr | NFS | 480,688,984 | 122,882,928 | eymir.grid.metu.edu.tr | eymir.grid.metu.edu.tr | 10.2 | 0 |
| g02.phy.bg.ac.yu (glite) | NFS | 118,169,880 | 23,038,880 | se.phy.bg.ac.yu | se.phy.bg.ac.yu | 0.04 | 0.09 |
| grid-ce.ii.edu.mk | NFS | 1,714,194,432 | 1,525,030,400 | grid-se.ii.edu.mk | grid-se.ii.edu.mk | 0.21 | 0 |
| grid001.ics.forth.gr | Local-SATA | 18,975,324 | 16,333,376 | Maxtor 6Y080M0 SATA, 150 MB/s, 7200 RPM, 8MB | grid002.ics.forth.gr | 0 | 0 |
| grid01.cu.edu.tr | Local-SATA | 10,080,488 | 6,884,432 | Segate Barracuda SATA 80 GB, 300 MB/s, 8 MB, 7200 RPM | grid02.cu.edu.tr | 0.63 | 0 |
| grid01.erciyes.edu.tr | Local-SATA | 10,080,488 | 2,068,864 | Segate Barracuda SATA 80 GB, 300 MB/s, 8 MB, 7200 RPM | grid02.erciyes.edu.tr | 0.63 | 0 |
| kalkan1.ulakbim.gov.tr (lcg) | NFS | 1,442,066,944 | 1,363,871,376 | torik1.ulakbim.gov.tr | torik1.ulakbim.gov.tr | 9.92 | 0.29 |
| kalkan2.ulakbim.gov.tr (glite) | NFS | 1,442,066,944 | 1,121,585,752 | torik1.ulakbim.gov.tr | torik1.ulakbim.gov.tr | 9.92 | 0.29 |
| node001.grid.auth.gr | NFS | 240,362,688 | 144,055,896 | node004.grid.auth.gr | node004.grid.auth.gr | 0.13 | 0.07 |
| paugrid1.pamukkale.edu.tr | Local-SATA | 58,586,904 | 55,578,036 | Segate Barracuda SATA 80 GB, 300 MB/s, 8 MB, 7200 RPM | paugrid2.pamukkale.edu.tr | 0.05 | 0 |
| testbed001.grid.ici.ro | NFS | 239,154,280 | 180,740,384 | testbed002.grid.ici.ro | testbed002.grid.ici.ro | 0.06 | 0 |
| wipp-ce.weizmann.ac.il | Local-ATA | 110,510,532 | 100,750,136 | West. Dig. Caviar ATA 120 GB, 100 MB/s, 2 MB, 7200 RPM | wipp-se.weizmann.ac.il | 0 | 0.14 |

From the second part of the script we have the results of the lmbench, which it was compiled in every worker node to take advantage of the full capabilities of it. The results were exported on a csv format and imported on an excel worksheet so that it can be more analyzed and compared among the different sites.

One of the most critical benchmarks is the numerical operations measurements. Most researches are based on number results and operations between those numbers. The time needed for those operations has a major role on the grid. The lmbench tests four types of numbers such ass integer, float, double, and 64bit integers and many kind of operations such as add, multiply, division and for integers an additional, the mod.

In the following table we can see the average of float and double number operations which are the result of running the lmbench on 3 different nodes on each site. The times are measured in nanoseconds and the smaller times are the better. Also we can see the best times with green color and the worst with red.

| Basic Float/Double Operations in nanoseconds (smaller is better) | | | | | | |
|---|---|---|---|---|---|---|
| SiteName | add | mul | div | add | mul | div |
| ce.hep.ntua.gr | 2.26 | 3.65 | 16.48 | 2.25 | 3.67 | 16.42 |
| ce.phy.bg.ac.yu (lcg) | 3.00 | 3.41 | 16.56 | 3.96 | 2.85 | 15.93 |
| ce.ulakbim.gov.tr | 1.88 | 2.64 | 16.26 | 1.88 | 2.63 | 16.26 |
| ce001.imbm.bas.bg | 1.78 | 2.50 | 15.43 | 1.78 | 2.50 | 15.43 |
| ce001.ipp.acad.bg (glite) | 2.00 | 2.67 | 15.11 | 2.01 | 2.68 | 15.14 |
| ce002.ipp.acad.bg (lcg) | 2.06 | 2.84 | 17.10 | 2.06 | 2.84 | 17.10 |
| ce01.afroditi.hellasgrid.gr | 1.76 | 2.35 | 13.30 | 1.76 | 2.36 | 13.31 |
| ce01.ariagni.hellasgrid.gr | 1.76 | 2.35 | 13.30 | 1.76 | 2.35 | 13.30 |
| ce01.athena.hellasgrid.gr (long) | 1.76 | 2.35 | 13.31 | 1.76 | 2.35 | 13.31 |
| ce01.athena.hellasgrid.gr (see) | 1.76 | 2.35 | 13.31 | 1.76 | 2.35 | 13.30 |
| ce01.isabella.grnet.gr | 2.54 | 10.89 | 15.54 | 2.85 | 5.14 | 15.52 |
| ce01.kallisto.hellasgrid.gr | 1.76 | 2.35 | 13.31 | 1.76 | 2.39 | 13.31 |
| ce01.marie.hellasgrid.gr | 1.77 | 2.36 | 13.30 | 1.76 | 2.35 | 13.30 |
| ce02.athena.hellasgrid.gr (long) | 1.76 | 2.35 | 13.30 | 1.76 | 2.35 | 13.30 |
| ce02.athena.hellasgrid.gr (see) | 1.76 | 2.35 | 13.30 | 1.76 | 2.35 | 13.30 |
| ce02.marie.hellasgrid.gr | 2.51 | 2.50 | 15.05 | 2.50 | 2.50 | 15.06 |
| ce101.grid.ucy.ac.cy | 1.54 | 1.54 | 9.30 | 1.55 | 1.54 | 9.29 |
| ctb31.gridctb.uoa.gr | 2.00 | 2.67 | 15.07 | 2.00 | 2.67 | 15.07 |
| cox01.grid.metu.edu.tr | 1.88 | 3.13 | 23.89 | 1.88 | 3.13 | 23.90 |
| g02.phy.bg.ac.yu (glite) | 4.68 | 5.79 | 16.64 | 4.69 | 5.62 | 16.69 |
| grid-ce.ii.edu.mk | 4.79 | 5.46 | 15.52 | 4.23 | 5.81 | 15.54 |
| grid001.ics.forth.gr | 2.15 | 2.87 | 16.21 | 2.15 | 2.87 | 16.21 |
| grid01.cu.edu.tr | 1.88 | 3.13 | 23.85 | 1.88 | 3.13 | 23.85 |
| grid01.erciyes.edu.tr | 1.88 | 3.13 | 23.85 | 1.88 | 3.14 | 23.85 |
| kalkan1.ulakbim.gov.tr (lcg) | 1.88 | 3.22 | 23.85 | 1.88 | 3.13 | 23.85 |
| kalkan2.ulakbim.gov.tr (glite) | 1.88 | 3.13 | 23.85 | 1.88 | 3.13 | 23.86 |
| node001.grid.auth.gr | 1.80 | 2.66 | 15.75 | 1.81 | 2.63 | 15.74 |
| testbed001.grid.ici.ro | 1.96 | 4.96 | 18.03 | 5.97 | 4.39 | 18.02 |
| wipp-ce.weizmann.ac.il | 2.78 | 4.42 | 32.02 | 2.78 | 4.42 | 32.00 |

The same tables with averages for integer and 64bit integer numbers are shown below.

**Basic Integer Operations in nanoseconds (smaller is better)**

| SiteName | bit | add | mul | div | mod |
|---|---|---|---|---|---|
| ce.hep.ntua.gr | 0.38 | 0.36 | 3.59 | 23.26 | 30.07 |
| ce.phy.bg.ac.yu (lcg) | 0.42 | 0.28 | 5.26 | 21.48 | 25.51 |
| ce.ulakbim.gov.tr | 0.19 | 0.19 | 5.30 | 21.79 | 24.26 |
| ce001.imbm.bas.bg | 0.18 | 0.18 | 5.03 | 20.68 | 23.01 |
| ce001.ipp.acad.bg (glite) | 0.33 | 0.33 | 3.34 | 20.72 | 27.80 |
| ce002.ipp.acad.bg (lcg) | 0.25 | 0.25 | 5.05 | 23.08 | 27.36 |
| ce01.afroditi.hellasgrid.gr | 0.29 | 0.29 | 2.94 | 18.24 | 24.63 |
| ce01.ariagni.hellasgrid.gr | 0.29 | 0.29 | 2.94 | 18.24 | 24.64 |
| ce01.athena.hellasgrid.gr (long) | 0.29 | 0.29 | 2.98 | 18.24 | 24.52 |
| ce01.athena.hellasgrid.gr (see) | 0.29 | 0.29 | 2.94 | 18.24 | 24.47 |
| ce01.isabella.grnet.gr | 0.27 | 0.22 | 5.18 | 22.19 | 25.14 |
| ce01.kallisto.hellasgrid.gr | 0.32 | 0.31 | 2.96 | 18.59 | 24.64 |
| ce01.marie.hellasgrid.gr | 0.29 | 0.31 | 2.94 | 18.24 | 24.54 |
| ce02.athena.hellasgrid.gr (long) | 0.29 | 0.29 | 2.94 | 18.24 | 24.48 |
| ce02.athena.hellasgrid.gr (see) | 0.30 | 0.29 | 2.95 | 18.24 | 24.67 |
| ce02.marie.hellasgrid.gr | 0.63 | 0.63 | 2.50 | 25.67 | 26.90 |
| ce101.grid.ucy.ac.cy | 0.39 | 0.39 | 1.16 | 15.83 | 16.21 |
| ctb31.gridctb.uoa.gr | 0.33 | 0.33 | 3.33 | 20.67 | 27.72 |
| cox01.grid.metu.edu.tr | 0.63 | 0.63 | 1.88 | 20.88 | 12.71 |
| g02.phy.bg.ac.yu (glite) | 0.43 | 0.27 | 5.13 | 21.66 | 24.18 |
| grid-ce.ii.edu.mk | 0.51 | 0.29 | 5.25 | 22.75 | 24.57 |
| grid001.ics.forth.gr | 0.36 | 0.36 | 3.58 | 22.20 | 29.99 |
| grid01.cu.edu.tr | 0.63 | 0.63 | 1.88 | 20.85 | 12.71 |
| grid01.erciyes.edu.tr | 0.63 | 0.63 | 1.88 | 20.84 | 12.71 |
| kalkan1.ulakbim.gov.tr (lcg) | 0.64 | 0.63 | 1.88 | 20.85 | 12.72 |
| kalkan2.ulakbim.gov.tr (glite) | 0.63 | 0.63 | 1.88 | 20.85 | 12.71 |
| node001.grid.auth.gr | 0.24 | 0.23 | 4.39 | 20.87 | 24.25 |
| testbed001.grid.ici.ro | 0.27 | 0.22 | 5.56 | 23.06 | 25.64 |
| wipp-ce.weizmann.ac.il | 0.82 | 0.82 | 3.91 | 34.44 | 38.94 |

**Basic Uint64 Operations in nanoseconds (smaller is better)**

| SiteName | bit | add | mul | div | mod |
|---|---|---|---|---|---|
| ce.hep.ntua.gr | 3.86 | 5.61 | 6.29 | 94.44 | 92.97 |
| ce.phy.bg.ac.yu (lcg) | 3.55 | 4.24 | 11.85 | 127.71 | 154.17 |
| ce.ulakbim.gov.tr | 3.24 | 3.41 | 13.00 | 80.82 | 95.61 |
| ce001.imbm.bas.bg | 3.08 | 3.25 | 12.36 | 76.68 | 90.76 |
| ce001.ipp.acad.bg (glite) | 3.22 | 5.58 | 5.46 | 79.15 | 70.70 |
| ce002.ipp.acad.bg (lcg) | 3.58 | 4.39 | 11.49 | 85.22 | 94.42 |
| ce01.afroditi.hellasgrid.gr | 3.13 | 4.91 | 4.85 | 67.56 | 62.02 |
| ce01.ariagni.hellasgrid.gr | 3.16 | 4.91 | 4.84 | 67.81 | 62.06 |
| ce01.athena.hellasgrid.gr (long) | 3.13 | 4.91 | 4.85 | 66.10 | 61.91 |
| ce01.athena.hellasgrid.gr (see) | 3.14 | 4.91 | 4.83 | 65.07 | 61.93 |
| ce01.isabella.grnet.gr | 3.43 | 3.74 | 12.91 | 91.98 | 108.10 |
| ce01.kallisto.hellasgrid.gr | 3.20 | 5.20 | 5.25 | 77.58 | 61.94 |
| ce01.marie.hellasgrid.gr | 2.98 | 5.11 | 4.82 | 67.94 | 62.18 |
| ce02.athena.hellasgrid.gr (long) | 2.99 | 4.91 | 4.81 | 69.68 | 62.07 |
| ce02.athena.hellasgrid.gr (see) | 3.10 | 4.92 | 4.90 | 65.04 | 62.03 |
| ce02.marie.hellasgrid.gr | 3.64 | 2.62 | 7.37 | 82.58 | 80.89 |
| ce101.grid.ucy.ac.cy | 2.57 | 1.72 | 3.33 | 48.02 | 45.70 |
| ctb31.gridctb.uoa.gr | 3.53 | 5.57 | 5.44 | 78.99 | 70.20 |
| cox01.grid.metu.edu.tr | 3.76 | 2.75 | 5.99 | 70.59 | 62.16 |
| g02.phy.bg.ac.yu (glite) | 3.67 | 4.07 | 11.46 | 115.52 | 126.76 |
| grid-ce.ii.edu.mk | 4.08 | 3.94 | 12.14 | 114.79 | 124.91 |
| grid001.ics.forth.gr | 3.69 | 5.99 | 5.86 | 84.92 | 75.57 |
| grid01.cu.edu.tr | 3.77 | 2.74 | 6.00 | 70.51 | 62.04 |
| grid01.erciyes.edu.tr | 3.77 | 2.74 | 5.99 | 70.51 | 62.03 |
| kalkan1.ulakbim.gov.tr (lcg) | 3.90 | 2.75 | 5.98 | 70.92 | 63.28 |
| kalkan2.ulakbim.gov.tr (glite) | 3.78 | 2.74 | 5.99 | 70.54 | 62.07 |
| node001.grid.auth.gr | 3.25 | 4.22 | 10.37 | 87.37 | 98.69 |
| testbed001.grid.ici.ro | 3.34 | 4.08 | 13.63 | 101.39 | 108.73 |
| wipp-ce.weizmann.ac.il | 5.75 | 4.67 | 8.96 | 127.22 | 123.45 |

One of the most important tasks of an operating system kernel is to manage processes and threads. A process is a program in execution and a thread is a CPU state stored within a process. A CPU context is either a process or a thread. A context switch occurs when processes are switched on the CPU. During this process the state of the old process is saved and the state of the new process is loaded. This means that context switching is pure overhead because the system does no work when a switch occurs.

The lmbench program measures context switching times. A number of processes are created, and they are linked by pipes into a ring. Each process writes to the next process down the ring a chunck of data. The benchmark times how long it takes to go from one process to another when the context switch is done by reading and writing data. Our benchmark gave results for the time it takes 2, 8, or 16 processes each using 0k, 16k, or 64k read and writes to make the results.

| Context switching - times in microseconds (smaller is better) | | | | | | | |
|---|---|---|---|---|---|---|---|
| SiteName | 2p/0K | 2p/16K | 2p/64K | 8p/16K | 8p/64K | 16p/16K | 16p/64K |
| ce.hep.ntua.gr | 2.52 | 2.44 | 2.63 | 3.47 | 5.67 | 3.77 | 12.13 |
| ce.phy.bg.ac.yu (lcg) | 3.95 | 3.69 | 14.50 | 6.58 | 65.60 | 17.50 | 83.60 |
| ce.ulakbim.gov.tr | 0.97 | 1.11 | 1.99 | 1.75 | 20.37 | 3.75 | 28.10 |
| ce001.imbm.bas.bg | 2.52 | 2.02 | 3.24 | 1.94 | 15.65 | 4.06 | 23.48 |
| ce001.ipp.acad.bg (glite) | 4.87 | 3.33 | 4.06 | 3.39 | 6.40 | 4.08 | 11.40 |
| ce002.ipp.acad.bg (lcg) | 1.88 | 2.24 | 2.60 | 2.62 | 12.71 | 3.95 | 22.40 |
| ce01.afroditi.hellasgrid.gr | 4.67 | 5.09 | 6.69 | 3.07 | 6.16 | 3.42 | 13.43 |
| ce01.ariagni.hellasgrid.gr | 2.70 | 3.19 | 3.71 | 3.78 | 5.86 | 3.75 | 15.03 |
| ce01.athena.hellasgrid.gr (long) | 3.99 | 6.79 | 6.94 | 3.26 | 5.51 | 3.92 | 11.29 |
| ce01.athena.hellasgrid.gr (see) | 4.74 | 6.43 | 8.32 | 4.77 | 5.46 | 4.24 | 12.23 |
| ce01.isabella.grnet.gr | 2.84 | 2.23 | 4.72 | 4.21 | 26.60 | 7.89 | 37.60 |
| ce01.kallisto.hellasgrid.gr | 2.45 | 2.87 | 2.33 | 4.27 | 6.30 | 3.47 | 13.23 |
| ce01.marie.hellasgrid.gr | 4.13 | 5.68 | 3.56 | 3.63 | 5.63 | 3.49 | 14.43 |
| ce02.athena.hellasgrid.gr (long) | 3.44 | 5.03 | 10.00 | 5.06 | 5.99 | 4.41 | 7.94 |
| ce02.athena.hellasgrid.gr (see) | 3.51 | 5.89 | 4.69 | 4.46 | 4.91 | 4.88 | 9.07 |
| ce02.marie.hellasgrid.gr | 1.75 | 2.17 | 10.07 | 3.51 | 131.33 | 23.00 | 138.77 |
| ce101.grid.ucy.ac.cy | 0.48 | 0.64 | 3.49 | 1.33 | 3.34 | 1.36 | 20.00 |
| ctb31.gridctb.uoa.gr | 4.77 | 4.85 | 2.69 | 2.72 | 2.93 | 2.80 | 5.15 |
| cox01.grid.metu.edu.tr | 4.06 | 2.20 | 1.51 | 3.57 | 2.21 | 2.32 | 1.62 |
| g02.phy.bg.ac.yu (glite) | 3.29 | 3.18 | 3.14 | 7.09 | 102.90 | 23.60 | 57.50 |
| grid-ce.ii.edu.mk | 2.24 | 2.32 | | 4.05 | 18.40 | 7.49 | 25.00 |
| grid001.ics.forth.gr | 5.13 | 5.44 | 5.27 | 3.69 | 5.28 | 4.12 | 12.80 |
| grid01.cu.edu.tr | 3.87 | 1.68 | 1.49 | 1.97 | 1.60 | 2.00 | 1.86 |
| grid01.erciyes.edu.tr | 4.06 | 4.00 | 3.67 | 2.61 | 2.21 | 2.80 | 1.68 |
| kalkan1.ulakbim.gov.tr (lcg) | 1.58 | 1.65 | 3.29 | 2.75 | 3.83 | 2.79 | 5.35 |
| kalkan2.ulakbim.gov.tr (glite) | 2.26 | 3.41 | 3.02 | 3.65 | 3.68 | 3.70 | 3.50 |
| node001.grid.auth.gr | 1.54 | 2.16 | 2.28 | 3.51 | 26.60 | 5.61 | 44.00 |
| testbed001.grid.ici.ro | 3.48 | 4.61 | 6.71 | 5.53 | 18.24 | 7.68 | 37.43 |
| wipp-ce.weizmann.ac.il | 2.95 | 3.47 | 3.81 | 9.54 | 72.08 | 21.35 | 96.67 |

In the first columns of "The File & VM system latencies" table we see the times that needed by the worker node to create and delete some sample files. This is a good benchmark if we consider of large file creation and deletion. Also we have the mmap latencies which is how fast a mapping can be made and unmade. This is useful because it is a fundemental part of processes that use SunOS style shared libraries (the libraries are mapped in at process start up time and unmapped at process exit). The lmbench maps in and unmaps the first \fIsize bytes of the file repeatedly and reports the average time for one mapping/unmapping.

| Site | 0K File create | 0K File delete | 10K File create | 10K File delete | Mmap Lat | Prot Fault | 100fd selct |
|---|---|---|---|---|---|---|---|
| ce.hep.ntua.gr | 27.17 | 22.27 | 317.27 | 491.57 | 9641 | 3.68 | 15.83 |
| ce.phy.bg.ac.yu (lcg) | 45.90 | 31.80 | 147.10 | 218.70 | 3364 | 5.18 | 22.50 |
| ce.ulakbim.gov.tr | 13.40 | 7.40 | 45.43 | 13.00 | 8920 | 2.74 | 2.84 |
| ce001.imbm.bas.bg | 15.10 | 10.74 | 73.37 | 31.87 | 3517 | 2.15 | 6.11 |
| ce001.ipp.acad.bg (glite) | 21.20 | 18.60 | 68.00 | 34.20 | 8809 | 2.83 | 12.40 |
| ce002.ipp.acad.bg (lcg) | 25.60 | 28.90 | 249.50 | 85.57 | 9702 | 2.95 | 6.11 |
| ce01.afroditi.hellasgrid.gr | 19.30 | 16.47 | 66.87 | 30.73 | 17267 |  | 11.57 |
| ce01.ariagni.hellasgrid.gr | 18.97 | 16.20 | 64.80 | 30.57 | 7990 | 2.54 | 11.37 |
| ce01.athena.hellasgrid.gr (long) | 18.23 | 15.60 | 63.10 | 29.47 | 16600 | 2.50 | 10.87 |
| ce01.athena.hellasgrid.gr (see) | 18.27 | 15.63 | 61.40 | 29.50 | 16567 | 2.50 | 10.87 |
| ce01.isabella.grnet.gr | 33.37 | 29.33 | 114.70 | 53.53 | 11222 |  | 20.20 |
| ce01.kallisto.hellasgrid.gr | 21.73 | 19.33 | 71.37 | 61.10 | 16600 | 2.74 | 12.17 |
| ce01.marie.hellasgrid.gr | 18.90 | 16.17 | 64.07 | 31.77 | 17000 | 2.58 | 11.40 |
| ce02.athena.hellasgrid.gr (long) | 18.20 | 15.70 | 60.50 | 29.50 | 16600 | 2.51 | 10.90 |
| ce02.athena.hellasgrid.gr (see) | 18.60 | 15.60 | 68.70 | 29.90 | 16600 | 2.51 | 10.80 |
| ce02.marie.hellasgrid.gr | 22.97 | 19.53 | 98.53 | 38.43 | 8323 | 3.15 | 6.74 |
| ce101.grid.ucy.ac.cy | 11.87 | 9.43 | 41.73 | 18.73 | 11414 | 2.1827 | 4.2 |
| ctb31.gridctb.uoa.gr | 21.40 | 18.80 | 65.20 | 32.80 | 17000 | 2.68 | 12.30 |
| cox01.grid.metu.edu.tr | 19.10 | 15.10 | 60.30 | 28.10 | 17000 | 3.88 | 11.80 |
| g02.phy.bg.ac.yu (glite) | 35.00 | 31.00 | 139.70 | 60.30 | 17800 | 5.57 | 22.00 |
| grid-ce.ii.edu.mk | 35.40 | 32.50 | 91.60 | 46.50 | 5581 | 4.31 | 18.60 |
| grid001.ics.forth.gr | 22.80 | 19.70 | 74.20 | 36.00 | 3070 | 1.99 | 13.40 |
| grid01.cu.edu.tr | 19.00 | 15.10 | 58.40 | 27.50 | 16900 | 2.82 | 11.90 |
| grid01.erciyes.edu.tr | 19.30 | 15.30 | 89.70 | 27.90 | 16900 | 2.86 | 11.90 |
| kalkan1.ulakbim.gov.tr (lcg) | 17.73 | 15.10 | 56.73 | 26.53 | 7515 |  | 11.93 |
| kalkan2.ulakbim.gov.tr (glite) | 17.87 | 15.10 | 57.40 | 26.10 | 8504 |  | 11.93 |
| node001.grid.auth.gr | 26.40 | 24.07 | 87.43 | 42.80 | 11067 | 3.31 | 18.30 |
| testbed001.grid.ici.ro | 82.60 | 80.33 | 274.07 | 951.53 | 17876 | 3.98 | 22.10 |
| wipp-ce.weizmann.ac.il | 26.60 | 21.93 | 95.17 | 41.30 | 21633 | 3.79 | 14.63 |

File & VM system latencies in microseconds (smaller is better)

In the next table in the first column we see the time needed to create a Unix pipe between two processes and move 50MB through the pipe in 64KB chunks. Most Unix systems implement the read/write system calls as a bcopy from/to kernel space to/from user space. Bcopy will use 2-3 times as much memory bandwidth. The write usually results in a cache line read and then a write back of the cache line at some later point.

*Local* Communication bandwidths in MB/s (bigger is better)

| Site | Pipe | AF Unix | TCP | File reread | Mmap reread | Bcopy (libc) | Bcopy (hand) | Mem read | Mem write |
|---|---|---|---|---|---|---|---|---|---|
| ce.hep.ntua.gr | 696 | 2883 | 1062 | 1797 | 3190 | 1084 | 1102 | 3237 | 1726 |
| ce.phy.bg.ac.yu (lcg) | 446 | 913 | 170 | 678 | 1071 | 431 | 318 | 834 | 473 |
| ce.ulakbim.gov.tr | 1395 | 2555 | 475 | 1909 | 2118 | 861 | 822 | 2037 | 1076 |
| ce001.imbm.bas.bg | 1254 | 2538 | 372 | 1640 | 1899 | 645 | 653 | 1871 | 922 |
| ce001.ipp.acad.bg (glite) | 648 | 2154 | 342 | 2231 | 4311 | 1139 | 1200 | 4308 | 1818 |
| ce002.ipp.acad.bg (lcg) | 1062 | 2261 | 223 | 1756 | 2372 | 814 | 821 | 2391 | 1137 |
| ce01.afroditi.hellasgrid.gr | 706 | 1927 | 382 | 1895 | 3246 | 900 | 876 | 3295 | 1323 |
| ce01.ariagni.hellasgrid.gr | 692 | 1805 | 395 | 2000 | 3769 | 994 | 984 | 3555 | 1397 |
| ce01.athena.hellasgrid.gr (long) | 717 | 1245 | 379 | 2155 | 3896 | 1120 | 1070 | 3886 | 1661 |
| ce01.athena.hellasgrid.gr (see) | 720 | 1300 | 376 | 2139 | 3873 | 1112 | 1060 | 3868 | 1677 |
| ce01.isabella.grnet.gr | 614 | 2807 | 328 | 970 | 1115 | 778 | 743 | 1662 | 1079 |
| ce01.kallisto.hellasgrid.gr | 763 | 1889 | 421 | 1886 | 3532 | 1017 | 961 | 3623 | 1552 |
| ce01.marie.hellasgrid.gr | 704 | 2125 | 608 | 2023 | 3500 | 1004 | 952 | 3742 | 1499 |
| ce02.athena.hellasgrid.gr (long) | 713 | 700 | 360 | 2113 | 3830 | 1085 | 1018 | 3840 | 1643 |
| ce02.athena.hellasgrid.gr (see) | 719 | 713 | 377 | 2123 | 3901 | 1093 | 1030 | 3882 | 1685 |
| ce02.marie.hellasgrid.gr | 794 | 683 | 166 | 357 | 565 | 298 | 295 | 519 | 451 |
| ce101.grid.ucy.ac.cy | 1564 | 1370 | 933 | 1355 | 2359 | 966 | 966 | 2027 | 1277 |
| ctb31.gridctb.uoa.gr | 814 | 2289 | 1516 | 2061 | 3975 | 1141 | 986 | 3895 | 1751 |
| cox01.grid.metu.edu.tr | 1160 | 1799 | 741 | 1383 | 2514 | 1202 | 1050 | 2519 | 1600 |
| g02.phy.bg.ac.yu (glite) | 356 | 990 | 118 | 514 | 589 | 270 | 223 | 588 | 392 |
| grid-ce.ii.edu.mk | 479 | 2022 | 323 | 31 | 37 | 753 | 889 | 1713 | 1083 |
| grid001.ics.forth.gr | 607 | 2133 | 499 | 2097 | 3681 | 1080 | 997 | 3715 | 1673 |
| grid01.cu.edu.tr | 1160 | 1212 | 558 | 1482 | 2564 | 1214 | 1077 | 2561 | 1654 |
| grid01.erciyes.edu.tr | 1163 | 2447 | 724 | 1451 | 2563 | 1210 | 1059 | 2565 | 1634 |
| kalkan1.ulakbim.gov.tr (lcg) | 1017 | 2502 | 806 | 1442 | 2564 | 1214 | 1054 | 2556 | 1627 |
| kalkan2.ulakbim.gov.tr (glite) | 788 | 2248 | 367 | 1054 | 1737 | 853 | 730 | 1715 | 1124 |
| node001.grid.auth.gr | 711 | 2979 | 614 | 1973 | 2419 | 1144 | 1111 | 2360 | 1445 |
| testbed001.grid.ici.ro | 504 | 2223 | 189 | 1046 | 1484 | 465 | 434 | 1046 | 646 |
| wipp-ce.weizmann.ac.il | 675 | 1699 | 533 | 1026 | 666 | 680 | 515 | 666 | 1050 |

Below we see the times for simple system calls. The null I/O benchmark measures the average of the times for a one-byte read from /dev/zero and a one-byte write to /dev/null. The stat benchmarks measures the time to invoke the stat system call on a temporary file. The open/close test measures the time to open a temporary file for reading and immediately close it. The fork, exec, and sh benchmarks measure three increasingly expensive forms of process creation: fork and exit, fork and execve, and fork and execlp of the shell with the new program as a command to the shell.

**Syscalls (System Calls) - times in microseconds (smaller is better)**

| Site | null call | null I/O | stat | open clos | slct TCP | fork proc | exec proc | sh proc |
|---|---|---|---|---|---|---|---|---|
| ce.hep.ntua.gr | 0.53 | 0.76 | 4.06 | 5.39 | 17.23 | 276 | 823 | 4019 |
| ce.phy.bg.ac.yu (lcg) | 0.60 | 0.84 | 5.14 | 6.74 | 25.70 | 529 | 2245 | 7909 |
| ce.ulakbim.gov.tr | 0.41 | 0.45 | 1.69 | 2.30 | 4.94 | 165 | 604 | 3154 |
| ce001.imbm.bas.bg | 0.38 | 0.47 | 2.12 | 2.80 | 7.57 | 147 | 490 | 2715 |
| ce001.ipp.acad.bg (glite) | 0.38 | 0.56 | 3.22 | 4.11 | 14.00 | 235 | 654 | 2987 |
| ce002.ipp.acad.bg (lcg) | 0.43 | 0.52 | 2.34 | 3.11 | 8.03 | 202 | 652 | 3228 |
| ce01.afroditi.hellasgrid.gr | 0.34 | 0.49 | 2.84 | 3.61 | 12.77 | 218 | 588 | 3019 |
| ce01.ariagni.hellasgrid.gr | 0.34 | 0.49 | 2.84 | 3.62 | 12.43 | 207 | 586 | 3088 |
| ce01.athena.hellasgrid.gr (long) | 0.34 | 0.49 | 2.82 | 3.61 | 12.30 | 208 | 577 | 2957 |
| ce01.athena.hellasgrid.gr (see) | 0.34 | 0.49 | 2.81 | 3.63 | 12.30 | 203 | 574 | 2850 |
| ce01.isabella.grnet.gr | 0.52 | 0.75 | 4.28 | 6.19 | 21.87 | 369 | 1102 | 5063 |
| ce01.kallisto.hellasgrid.gr | 0.36 | 0.53 | 3.10 | 4.15 | 12.43 | 231 | 639 | 3314 |
| ce01.marie.hellasgrid.gr | 0.34 | 0.49 | 2.82 | 3.62 | 12.40 | 208 | 578 | 2995 |
| ce02.athena.hellasgrid.gr (long) | 0.34 | 0.49 | 2.82 | 3.61 | 12.30 | 206 | 585 | 2997 |
| ce02.athena.hellasgrid.gr (see) | 0.33 | 0.49 | 2.82 | 3.62 | 12.30 | 208 | 582 | 3009 |
| ce02.marie.hellasgrid.gr | 0.19 | 0.36 | 3.68 | 4.60 | 28.07 | 209 | 1094 | 5096 |
| ce101.grid.ucy.ac.cy | 0.12 | 0.23 | 1.86 | 2.40 | 10.93 | 149 | 480 | 2088 |
| ctb31.gridctb.uoa.gr | 0.36 | 0.53 | 3.23 | 4.12 | 14.00 | 218 | 615 | 2858 |
| cox01.grid.metu.edu.tr | 0.38 | 0.55 | 2.77 | 3.61 | 14.00 | 194 | 1106 | 3602 |
| g02.phy.bg.ac.yu (glite) | 0.56 | 0.86 | 5.08 | 6.66 | 25.60 | 570 | 2066 | 8452 |
| grid-ce.ii.edu.mk | 0.58 | 0.88 | 5.08 | 6.97 | 25.70 | 415 | 1367 | 6023 |
| grid001.ics.forth.gr | 0.41 | 0.60 | 3.47 | 4.43 | 15.10 | 208 | 555 | 2896 |
| grid01.cu.edu.tr | 0.38 | 0.55 | 2.78 | 3.63 | 14.00 | 196 | 610 | 2998 |
| grid01.erciyes.edu.tr | 0.38 | 0.55 | 2.76 | 3.64 | 14.00 | 205 | 615 | 3088 |
| kalkan1.ulakbim.gov.tr (lcg) | 0.39 | 0.59 | 2.37 | 3.19 | 15.20 | 132 | 388 | 2802 |
| kalkan2.ulakbim.gov.tr (glite) | 0.38 | 0.61 | 5.63 | 4.10 | 14.50 | 123 | 387 | 2536 |
| node001.grid.auth.gr | 0.46 | 0.68 | 3.75 | 5.07 | 19.90 | 257 | 805 | 3901 |
| testbed001.grid.ici.ro | 0.60 | 0.80 | 4.36 | 6.29 | 23.20 | 300 | 1202 | 4859 |
| wipp-ce.weizmann.ac.il | 0.37 | 0.57 | 3.57 | 4.53 | 17.43 | 300 | 990 | 4471 |

In the next table we see measures about memory read latency for varying memory sizes and strides. The results are reported in nanoseconds per load. The entire memory hierarchy is measured, including onboard cache latency and size, external cache latency and size, main memory latency, and TLB miss latency.

| Memory latencies in nanoseconds (smaller is better) | | | | |
|---|---|---|---|---|
| Site | L1$ | L2$ | Main Mem | Rand Mem |
| ce.hep.ntua.gr | 1.538 | 10.529 | 43.6 | 196.6 |
| ce.phy.bg.ac.yu (lcg) | 0.862 | 14.600 | 242.8 | 318.3 |
| ce.ulakbim.gov.tr | 0.751 | 6.883 | 114.7 | 216.4 |
| ce001.imbm.bas.bg | 0.920 | 7.455 | 104.3 | 201.5 |
| ce001.ipp.acad.bg (glite) | 1.339 | 9.359 | 37.1 | 198.7 |
| ce002.ipp.acad.bg (lcg) | 1.008 | 8.233 | 82.4 | 199.7 |
| ce01.afroditi.hellasgrid.gr | 1.177 | 8.337 | 58.2 | 288.9 |
| ce01.ariagni.hellasgrid.gr | 1.177 | 8.342 | 50.8 | 216.4 |
| ce01.athena.hellasgrid.gr (long) | 1.177 | 8.230 | 48.9 | 233.5 |
| ce01.athena.hellasgrid.gr (see) | 1.178 | 8.265 | 49.1 | 233.9 |
| ce01.isabella.grnet.gr | 0.996 | 10.931 | 127.0 | 278.7 |
| ce01.kallisto.hellasgrid.gr | 1.177 | 8.248 | 49.4 | 231.1 |
| ce01.marie.hellasgrid.gr | 1.179 | 8.291 | 37.0 | 229.5 |
| ce02.athena.hellasgrid.gr (long) | 1.178 | 8.250 | 49.5 | 242.7 |
| ce02.athena.hellasgrid.gr (see) | 1.177 | 8.209 | 48.9 | 240.1 |
| ce02.marie.hellasgrid.gr | 1.876 | 12.500 | 201.3 | 408.7 |
| ce101.grid.ucy.ac.cy | 1.157 | 4.894 | 84.5 | 137.1 |
| ctb31.gridctb.uoa.gr | 1.334 | 9.324 | 46.8 | 273.4 |
| cox01.grid.metu.edu.tr | 1.882 | 8.814 | 135.4 | 198.4 |
| g02.phy.bg.ac.yu (glite) | 0.893 | 14.400 | 363.3 | 695.2 |
| grid-ce.ii.edu.mk | 0.780 | 15.600 | 114.5 | 248.7 |
| grid001.ics.forth.gr | 1.434 | 10.000 | 51.1 | 188.9 |
| grid01.cu.edu.tr | 1.880 | 8.814 | 135.4 | 157.7 |
| grid01.erciyes.edu.tr | 1.880 | 8.814 | 135.4 | 182.3 |
| node001.grid.auth.gr | 0.715 | 6.519 | 108.6 | 211.1 |
| testbed001.grid.ici.ro | 1.255 | 10.889 | 132.0 | 265.4 |
| wipp-ce.weizmann.ac.il | 1.911 | 8.362 | 137.8 | 279.2 |